\documentclass[pra, reprint]{revtex4-1}
\pdfoutput=1
\usepackage{amsmath}
\usepackage{amsfonts}
\usepackage{graphicx}
\usepackage{physics}
\usepackage{multirow}
\usepackage{siunitx}

\begin{document}
\author{W. Y. Kon}
\affiliation{School of Physical and Mathematical Sciences, Nanyang Technological University, Singapore}
\affiliation{Department of Physics and Astronomy, Rice University, Houston, Texas 77005, USA}
 
\author{J. A. Aman}
\affiliation{Department of Physics and Astronomy, Rice University, Houston, Texas 77005, USA}

\author{J. C. Hill}
\affiliation{Department of Physics and Astronomy, Rice University, Houston, Texas 77005, USA}
 
\author{T. C. Killian}
\affiliation{Department of Physics and Astronomy, Rice University, Houston, Texas 77005, USA}

\author{Kaden R. A. Hazzard}
\email{kaden@rice.edu}
\affiliation{Department of Physics and Astronomy, Rice University, Houston, Texas 77005, USA}

\title{High-intensity two-frequency photoassociation spectroscopy of a weakly bound molecular state: theory and experiment}
\date{\today}

\begin{abstract}
	We investigate two-frequency photoassociation of a weakly bound molecular state, focusing on a regime where the AC Stark shift is comparable to the halo-state energy. In this ``high-intensity'' regime, we observe features absent in low-intensity two-frequency photoassociation. We experimentally measure the spectra of $^{86}$Sr atoms coupled to the least bound state of the $^{86}$Sr$_2$ ground electronic channel through an intermediate electronically excited molecular state. We compare the spectra to a simple three-level model that includes a two-frequency drive on each leg of the transition.
	With numerical solution of the time-dependent Schr{\"o}dinger equation, we show that this model accurately captures (1) the existence of experimentally observed satellite peaks that arise from nonlinear processes, (2) the locations of the two-photon peak in the spectrum, including AC Stark shifts, and (3) in some cases, spectral lineshapes.  To better understand these numerical results, we develop an approximate treatment of this model, based on Floquet and perturbation theory, that gives simple formulas that accurately capture the halo-state energies.  
	We expect these expressions to be valuable tools to analyze and guide future two-frequency photoassociation experiments. 
\end{abstract}

\maketitle

\section{Introduction}

The coherent coupling of atoms to bound dimers using light is a fundamental tool in ultracold physics. It provides a sensitive probe of interatomic potentials~\cite{Lett1993,Miller1993}, can be used to control the interaction strength with an optical Feshbach resonance~\cite{Fedichev1996,Theis2004,Chin2010}, and is a key step in creating ultracold ground state molecules~\cite{Jones2006,Vanhaecke2004,Kim2004,Bohn2017}.

Recent experiments~\cite{Jim2018} have spectroscopically probed a weakly-bound halo state in ultracold alkaline-earth atoms. Halo state molecules are molecules with a large spatial extent, with most of the wavefunction residing in the classically forbidden region~\cite{Riisager2000}. Ultracold alkaline-earth atoms are an exciting form of quantum matter~\cite{Stellmer2014}, offering extremely long-lived electronic clock states~\cite{Ludlow2015}, a large nuclear spin degeneracy in the fermionic isotopes (up to $N=10$ states for $^{87}$Sr), and an SU($N$) symmetry that can stabilize quantum fluctuations~\cite{Cazalilla2014,Gorshkov2010}.

\begin{figure*}
	\includegraphics[width=\textwidth]{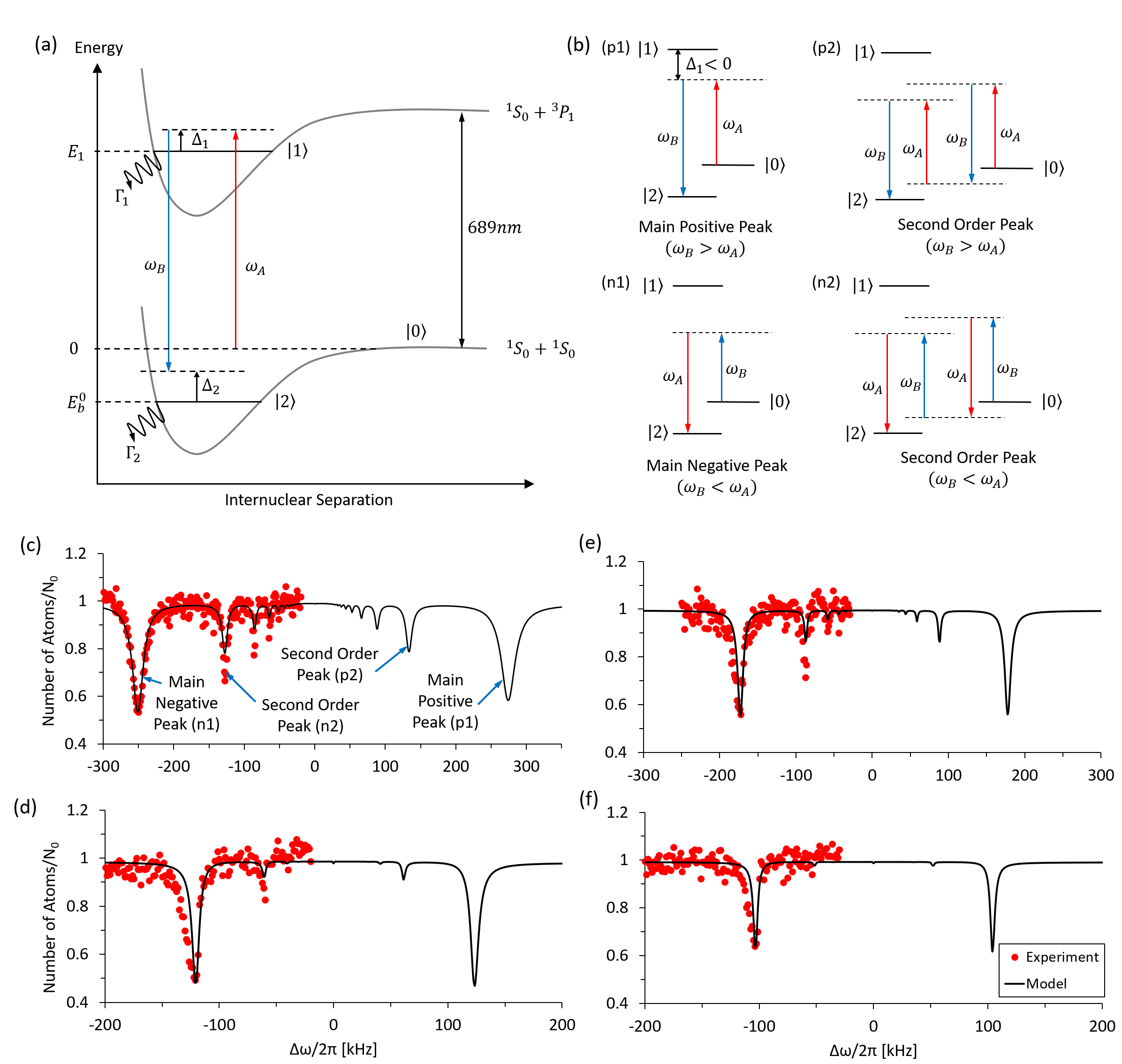}
	\caption{\label{SystemPic}(a) A schematic of the two-frequency PA process. $E_1$ is the intermediate state energy, $\Delta_1$ is the detuning of laser $\omega_A$ from the $0\rightarrow 1$ transition, and $\Delta_2$ is the two-photon detuning with respect to the $0\rightarrow 2$ transition. (b) Possible resonance processes at different PA laser frequencies, $\omega_A$ and $\omega_B$, that correspond to different peaks observed in experimental PA spectra, as labeled in (c). The second-order peaks only become possible if we assume both lasers drive each leg of the transition. (c-f) Atom loss spectrum obtained from PA experiment (circles) and numerical solutions of the three-level model (solid line) as a function of PA laser frequency difference, $\Delta\omega=\omega_B-\omega_A$, (c) with detuning $\Delta_1/2\pi=$ \SI{-1.5}{\mega\hertz} and intensity $I= $ \SI{0.66}{\watt\per\centi\meter\squared}, (d) with detuning $\Delta_1/2\pi=$ \SI{-1.5}{\mega\hertz} and intensity $I= $ \SI{0.17}{\watt\per\centi\meter\squared}, (e) with detuning $\Delta_1/2\pi=$ \SI{-3}{\mega\hertz} and intensity $I= $ \SI{0.66}{\watt\per\centi\meter\squared}, and (f) with detuning $\Delta_1/2\pi=$ \SI{-3}{\mega\hertz} and intensity $I= $ \SI{0.18}{\watt\per\centi\meter\squared}. Note that the value of $k$ [Eq.~\eqref{kdef}] used in the three-level model is different for each spectrum. Trap parameters, densities and other experimental conditions are provided in Ref.~\cite{Jim2018}.}
\end{figure*}

Accessing halo states in alkaline-earth atoms requires different tools than in alkali atoms, as the former's closed electronic structure precludes the existence of magnetic Feshbach resonances~\cite{Chin2010} and the formation of weakly bound molecular states through magnetoassociation~\cite{Kohler2006}, and it forbids one-photon processes from coupling a well-separated pair of atoms to the halo dimer~\cite{Chin2005}. Consequently, Ref.~\cite{Jim2018} used two-photon photoassociation (PA) through an intermediate molecular bound state in an excited electronic channel, following techniques in Refs.~\cite{Jones2006,Escobar2008,Abraham1995}, as illustrated in Fig.~\ref{SystemPic}(a). If the frequencies $\omega_A$ and $\omega_B$ of the two lasers are sufficiently different, only one laser will be relevant for driving each leg of the two-photon transition, corresponding to the commonly used $\Lambda$-model~\cite{Yoo1985,Bohn1996}. When the intermediate state detuning is much larger than the Rabi frequency of each leg of the transition, the $\Lambda$-model predicts a single photoassociation peak at $\omega_B-\omega_A=-E_b$, where $E_b<0$ is the halo-state energy (including light shift). (Here and throughout, $\hbar=1$.)

However, this description is not always accurate. One such case is when the magnitude of the unperturbed halo-state energy, $E_b^0$, is not large relative to the detuning from the intermediate state. In this case, the PA lasers are similar enough in frequency that both are relevant for each leg of the transition. This two-frequency drive for each leg of the transition leads to additional AC Stark shifts, as we will discuss in Sec.~\ref{SecResPos}. 

A second case where treating each leg with a single-frequency drive is inadequate is when $E_b^0$ (which we will also refer to as the unperturbed binding energy) is not much larger than the AC Stark shift. In this regime, additional peaks appear in the PA spectra at approximately $E_b/2$, $E_b/3$, $\ldots$, as shown in Fig.~\ref{SystemPic}(c). This occurs at high intensities and/or low intermediate-state detunings, and we will refer to this as the ``high-intensity'' regime throughout this paper. We will discuss the origin of this effect in detail in Sec.~\ref{MultiphotonSec}. As per Fig.~\ref{SystemPic}(c), we will refer to the most red-detuned line in the spectrum as the main negative peak, the most blue-detuned line as the main positive peak, and the additional peaks as higher order peaks.

In this paper, we measure spectra in the high-intensity regime and develop a theory that accounts for the effects of the two-frequency drive when the molecular binding energy is small and excitation laser intensity is high. The model neglects motional degrees of freedom for the initial state of two free atoms and considers only three levels: a dimer in the ground electronic state, the intermediate dimer state in the excited electronic potential, and two well-separated atoms. We numerically solve this model and analyze many of its features with analytic approximations. We also briefly consider extensions of this three-level model that add extra levels to capture some fine details of the AC Stark shift at large detuning.

We show that the three-level model captures key experimentally observed features despite its simplicity. Numerical simulations show that it accurately predicts the appearance of additional spectral lines in the high-intensity regime, and that these correspond to higher order non-linear processes. It accurately captures the location of the spectral peaks, including AC Stark shifts from both lasers when intermediate state detuning, $\Delta_1$, is comparable to $E_b^0$. Furthermore, when the intensity of the lasers is not too small, the spectral lineshapes of individual lines are reasonably well-captured. This indicates that effects missing from this theory -- the motion of the atoms~\cite{Bohn1996}, interaction shifts due to molecules and atoms scattering off of other molecules and atoms~\cite{Wynar2000}, and the varying density within the trap~\cite{Escobar2008} -- are less relevant in the regime of interest here.

We also develop a simplified analytic theory based on a Floquet treatment of this model, similar to Ref.~\cite{Ho1984,Wang1985}, which captures the binding energy of the spectral peaks, including their dependence on laser frequency and intensity. The simple expression for binding energy is given in Eq.~\eqref{FloqFinalEq}. This will allow future experiments to easily predict and analyze these features.

The experimental system used here is $^{86}$Sr, which has a weakly bound halo state in its ground electronic channel. The unperturbed binding energy of this halo state was accurately determined in a companion paper~\cite{Jim2018}. Otherwise, the only previous PA experiments with $^{86}$Sr have been single-photon PA to excited electronic states~\cite{Mickelson2005,Borkowski2014,Reschovsky2018}. Photoassociation of Sr in general has been extensively studied because of interest in its scattering properties~\cite{Mickelson2005,Yasuda2006,Escobar2008}, optical Feshbach resonances~\cite{Blatt2011,Yan2013,Nicholson2015}, excited state molecular potentials~\cite{Zelevinsky2006,Borkowski2014,Reschovsky2018}, quantum chemistry~\cite{McGuyer2013,McGuyer2014,McGuyer2015,Kondov2018,Majewska2018}, ultracold chemistry~\cite{McDonald2016,Yan2013}, and the formation of ground state molecules~\cite{Stellmer2012,Ciamei2017,Koch2008}, which has important applications, for example, in precision measurements~\cite{Zelevinsky2008}. We perform two-frequency photoassociation~\cite{Abraham1995,Bohn1999,Jones2006} of a ground state Sr molecule~\cite{Stellmer2012,Ciamei2017,Koch2008,Jim2018}, similar to experiments in Yb~\cite{Kitagawa2008} and Ca~\cite{Pachomow2017}.

The Sr system has been extensively studied with ab initio methods that predict photoassociation rates from first principles~\cite{Moszynski2012}. We emphasize that here we focus on an isolated few-level system and the new photassociation phenomena that arise at high laser intensity where the Rabi frequency is not small compared to Sr$_2$ halo state binding energy. We note that this is still a very low laser intensity in another sense, in that the Rabi frequency remains small compared to the detuning from other molecular, rotational and vibrational states not part of the isolated few-level system considered here. Rather than calculate Frank-Condon factors and dipole transition matrix elements from an ab initio approach~\cite{Moszynski2012,Skomorowski2012}, we take these as phenomenological fit parameters in our theoretical descriptions. This is in the same spirit as typical applications of the Bohn-Julienne description to photoassociation spectra~\cite{Bohn1999,Jones2006}, where only a few parameters -- which in principle are related to microscopic wavefunctions -- are important, and these are determined from experiment. Once these parameters are fixed, this theory is expected to be accurate when collisional energies and Rabi frequencies are low compared to the detuning to other energy states (ie where the system can still be considered an isolated few-level system). This formalism has been applied successfully in many settings~\cite{Lisdat2002,Araujo2003}, including systems with complicated electronic structures such as Ca~\cite{Pachomow2017}, Sr~\cite{Zelevinsky2006,Escobar2008,Jim2018}, and Yb~\cite{Borkowski2009}.

In Sec.~\ref{ExpSec}, we describe the PA experiment and the key features observed from the spectra. We introduce and develop the three-level model in Sec.~\ref{ModelSec}. In Sec.~\ref{SecResPos}, we evaluate the model's ability to replicate the binding energy obtained in experiment and introduce an analytic treatment of the model using Floquet and perturbation theory. Sec.~\ref{LineShapeSec} examines the effectiveness of the three-level model in predicting spectral linshapes. Sec.~\ref{MultiphotonSec} explores higher order processes that become possible when both lasers driving each leg of the transition are relevant, and describes additional peaks observed in experiment.

\section{Experiment}
\label{ExpSec}

We perform two-frequency PA spectroscopy of $^{86}$Sr atoms, where the atoms are coupled to a halo state on the molecular potential of the $^{86}$Sr$_2$ ground electronic state through an intermediate molecular bound state. Two PA lasers with frequencies $\omega_A$ and $\omega_B$, intensities $I_A$ and $I_B$, and the same linear polarization are applied to ultracold $^{86}$Sr in an optical dipole trap, as illustrated in Fig.~\ref{SystemPic}(a). The temperature in the trap is typically \SI{300}{\nano\kelvin}. We use equal PA laser intensities in the experiment, $I_A=I_B=I$. By scanning the frequency of laser B, we obtain a spectrum of atom loss as a function of the frequency difference of PA lasers, i.e. $\Delta\omega=\omega_B-\omega_A$, in the regime $\Delta\omega<0$. When the two-photon process is resonant with the binding energy, $\Delta\omega=E_b$, there is significant atom loss through inelastic processes acting on the bound halo state and intermediate excited molecular state, and we expect a peak to be present in the atom loss spectrum. From the spectral peaks, we extract the binding energy by fitting the lineshape to a convolution of the asymmetric loss function described in Ref.~\cite{Jim2018} with a Lorentzian, which provides good fit to the lineshape for all intensities. We measure spectra for different intermediate-state detunings, $\Delta_1=\omega_A-E_1$, and PA laser intensities to determine the dependence of light shifts on these parameters. More details of the experimental setup can be found in Ref.~\cite{Jim2018}, where low-intensity spectra are reported and the unperturbed binding energy, $E_b^0$, was determined to be -83.0 $\pm$ \SI{0.3}{\kilo\hertz}.

In high-intensity measurements reported in the present paper, we observe features absent in the $\Lambda$-model where only a single laser drives each leg of the PA process. Most notable are additional peaks near $\Delta\omega = \pm E_b/n$ for integer $n$, as shown in the spectra in Fig.~\ref{SystemPic}(c-e). In Sec.~\ref{MultiphotonSec}, we will explain these peaks in terms of non-linear processes illustrated in Fig.~\ref{SystemPic}(b) that become possible with the two-frequency drive. These additional peaks are more apparent at higher PA laser intensity and lower detuning.

The peaks we observe in the experiment also experience strong light shifts, and understanding these is complicated by both lasers driving each leg of the transition. We gain insight into the challenges that the two-frequency drive presents for predicting the AC Stark shift by considering two regimes. The first limit is when $\omega_A$ and $\omega_B$ are far apart, corresponding to large binding energy. The two lasers oscillate at very different frequencies and thus we do not expect the interference term to contribute. In this case, the AC Stark effects from each laser add independently, giving a contribution proportional to $\frac{I_A}{4(\Delta_1-\Delta\omega)}+\frac{I_B}{4\Delta_1}$ for $\Delta\omega>0$, a formula that has been suggested and used in Refs.~\cite{Kitagawa2008,Wynar2000}. The second limit is when $\omega_A$ and $\omega_B$ are nearly equal and the electric fields $\vec{E}_A$ and $\vec{E}_b$ add constructively, where the total electric field amplitude is $\vec{E}=\vec{E}_A+\vec{E}_B$. In this case, one obtains a Stark shift proportional to $\frac{I_A+I_B+2\sqrt{I_AI_B}}{4\Delta_1}$. The experiments reported in this paper were performed at an intermediate regime to these limits. Section~\ref{SecNumFit} will present results from numerical simulations that describe the shifts in this intermediate regime, and Section~\ref{FloqMain} derives analytic expressions capable of capturing the behavior in relevant limits.

\section{Three-Level Model}
\label{ModelSec}

We model the PA experiment using a three-level system with $\ket{0}$ representing two well-separated $^{86}$Sr atoms in the $^{1}$S$_0$ state, $\ket{1}$ representing the second least-bound vibrational state of the $0_u^+$ molecular potential, and $\ket{2}$ representing the halo state, as shown in Fig.~\ref{SystemPic}(a). Two lasers with frequencies $\omega_A$ and $\omega_B$, corresponding to the PA lasers, are included to drive the two-frequency PA process. Unlike the $\Lambda$-model where each laser will only drive one transition, our model includes additional couplings where laser A drives the transition $1\rightarrow 2$ and laser B drives the transition $0\rightarrow 1$. We include decay terms to describe atom loss in the system, with $\Gamma_1$ representing atom loss from the intermediate bound state due to natural decay and $\Gamma_2$ representing loss from the halo state due to collisions with background atoms. The Hamiltonian is thus
\begin{equation}
\label{OriginalH}
\begin{split}
\hat{H}=&\left[\Omega_{A,01}\cos(\omega_At)+\Omega_{B,01}\cos(\omega_Bt)\right]\dyad{0}{1}+\text{h.c.}\\
&+\left[\Omega_{B,12}\cos(\omega_Bt)+\Omega_{A,12}\cos(\omega_A t)\right]\dyad{1}{2}+\text{h.c.}\\
&+\left(E_1-i\frac{\Gamma_1}{2}\right)\dyad{1}+\left(E_b^0-i\frac{\Gamma_2}{2}\right)\dyad{2}
\end{split}
\end{equation}
where $\Omega_{L,T}$ is the coupling from laser $L$ driving transition $T$. Using the rotating wave approximation and a suitable basis change, we obtain
\begin{equation}
\label{SimulationH}
\begin{split}
\hat{H}=&\left(\frac{\Omega_{A,01}}{2}+\frac{\Omega_{B,01}}{2}e^{i\Delta\omega t}\right)\dyad{0}{1}+\text{h.c.}\\
&+\left(\frac{\Omega_{B,12}}{2}+\frac{\Omega_{A,12}}{2}e^{i\Delta\omega t}\right)\dyad{1}{2}+\text{h.c.}\\
&+\left(-\Delta_1-i\frac{\Gamma_1}{2}\right)\dyad{1}+\left(-\Delta_2-i\frac{\Gamma_2}{2}\right)\dyad{2}
\end{split}
\end{equation}
where $\Delta_2 = -\Delta\omega-E_b^0$. In the experiment, both laser intensities are equal, and thus we expect $\Omega_{B,12}=\Omega_{A,12}$ and $\Omega_{B,01}=\Omega_{A,01}$. However, the couplings for the two transitions are different since the corresponding matrix elements are different, and they can be characterised by the parameter
\begin{equation}
x=\frac{\Omega_{B,12}}{\Omega_{A,01}}.
\end{equation}
From the results of Ref.~\cite{Jim2018}, we know that for modeling our experiment, $x$ is large ($x\gg1$). Since the Rabi frequencies depend on the laser electric field as $\Omega\propto\vec{E}$, we have
\begin{equation}
\label{kdef}
\Omega_{B,12}=2\pi k\sqrt{I_B} 
\end{equation}
for some constant $k$ that we will determine through fits with experimental data. 

Our model neglects motional effects. However, photoassociation is a scattering process, involving an initial state of two free atoms, and thus the collision amplitude will depend on energy. As we will show in Sec.~\ref{LineShapeSec}, this limits our ability to accurately predict lineshapes in the low intensity regime in which spectra are broadened by the distribution of relative kinetic energies for initial scattering states.

Nevertheless, numerically simulated spectra using the three-level model do reproduce many features observed in the experiments, such as additional peaks, AC Stark shifts for a wide range of intensities and detuning, and high-intensity lineshapes, as shown in Fig.~\ref{SystemPic}(c-f).

The model produces an asymmetry between the $\Delta\omega>0$ and $\Delta\omega<0$ regime in the spectra. To understand this, note that the AC Stark shift is dominated by laser coupling on the $1\rightarrow 2$ transition, and it largely depends on the one-photon detunings of both lasers from this resonance. To obtain the atom loss spectra, we fix $\omega_A$ and scan $\omega_B$. For the conditions of data in Fig.~\ref{SystemPic}(c)-(f), as $\Delta\omega=\omega_B-\omega_A$ increases from large negative to large positive, the single-photon detuning of laser B decreases and its contribution to the AC Stark shift increases. The contribution from laser A remains approximately constant.

\begin{figure*}
	\includegraphics[width=0.9\textwidth]{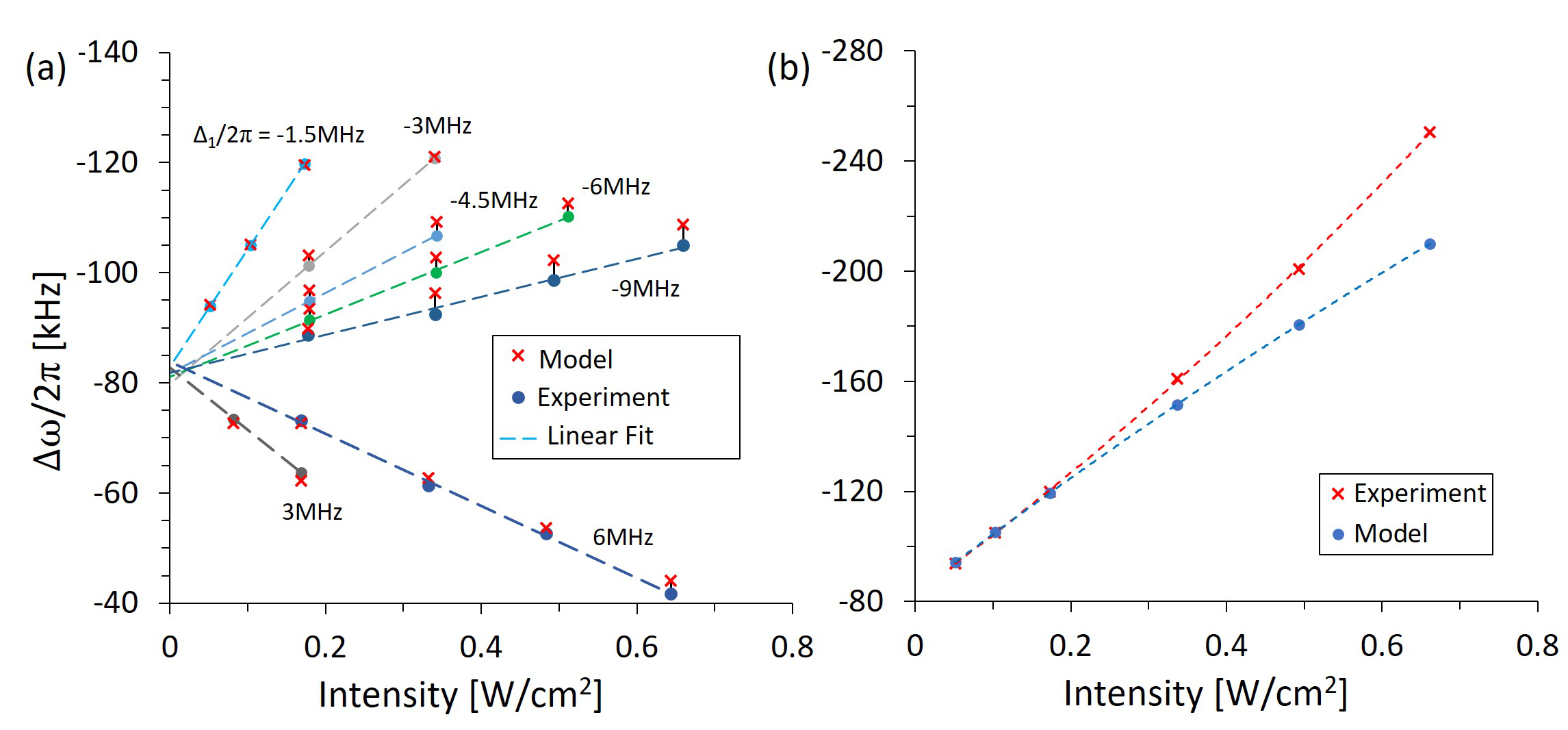}
	\caption{\label{ExpData}(a) Difference in laser frequency, $\Delta\omega$, at peak atom loss, which we identify as the energy difference between state $\ket{0}$ and $\ket{2}$. This corresponds to the energy (unperturbed energy + AC Stark shift) of the halo ground-state molecule. The frequency at peak loss as a function of intensity is obtained from numerical simulation of the three-level model, given by Eq.~\eqref{SimulationH}, and the PA experiment at various detunings, $\Delta_1$. We obtain a global $k=$ \SI[mode=text]{850}{kHz(W/cm^2)^{-1/2}} by fitting the $\Delta_1/2\pi=$ \SI{-1.5}{\mega\hertz} data points. Note that the large detuning data is not used to determine $k$, but it contributes to the analysis in Fig.~\ref{Floquet4Level}. (b) Same plot of the $\Delta_1/2\pi=$ \SI{-1.5}{\mega\hertz} data as for (a), except that higher intensity data is included, for which the three-level model predictions deviate from experimental results. Quadratic fits are plotted to guide the eye.}
\end{figure*}

\section{Binding Energy}
\label{SecResPos}

\subsection{Numerical solution of three-level model}
\label{SecNumFit}

To determine the model parameters $k$, $x$, $E_b^0$, $\Gamma_1$, and $\Gamma_2$, we numerically solve the time-dependent Schr{\"o}dinger equation with the Hamiltonian in Eq.~\eqref{SimulationH} with initial state $\ket{0}$. The square of the magnitude of the coefficient of $\ket{0}$ in the wavefunction after some time $\tau$ (duration the PA lasers are on) gives the fraction of atoms remaining in the optical trap, and this calculation can be repeated at different $\Delta\omega$ values to obtain an atom loss spectrum, as shown in Fig.~\ref{SystemPic}(c-f). A normalization parameter, $N_0$, is introduced to match the calculated number of atoms remaining to that of the experiment, and its value differs between plots at different intensities and intermediate state detunings. We identify the laser frequency difference $\Delta\omega$ of the atom loss peak as the energy difference between $\ket{0}$ and $\ket{2}$, which corresponds to the molecular binding energy, and we adjust the parameters to match simulation with experiment. We find that $x$, $\Gamma_1$, and $\Gamma_2$ have little impact on the location of the main negative peak [labeled n1 in Fig.~\ref{SystemPic}(c)]. Therefore, we fix $x=$ 40, $\Gamma_1 =$ \SI{1.1e6}{\per\second} and $\Gamma_2 =$ \SI{1.6e4}{\per\second} for our analysis, which can replicate roughly the main peak's lineshape for the $I=$ \SI{0.66}{\watt/\centi\metre\squared} and $\Delta_1 =$ \SI{-1.5}{\mega\hertz} spectrum in Fig.~\ref{SystemPic}(c). More accurate ways of determining the values of $x$, $\Gamma_1$, and $\Gamma_2$ are examined in Sec.~\ref{LineShapeSec}. 

At the range of intensities and detunings shown in Fig.~\ref{ExpData}(a), we observe that the three-level model generally yields a linear relation between binding energy and intensity. Hence, we obtain a value for the unperturbed binding energy of \SI{-82.9}{\kilo\hertz} by linear extrapolation of the low-intensity \SI{-1.5}{\mega\hertz} detuning data shown in Fig.~\ref{ExpData}(a). This value is consistent with the more accurate determination given in Ref.~\cite{Jim2018}, which additionally takes into account the effects of atomic densities, trapping lasers and initial scattering state. A global value of $k=$ \SI[mode=text]{850}{kHz(W/cm^2)^{-1/2}} is then determined by minimizing the least square distance between the model's predicted binding energy and the same \SI{-1.5}{\mega\hertz} detuning data.

With just two effective fitting parameters, $k$ and $E_b^0$ ($x$, $\Gamma_1$, and $\Gamma_2$ have a negligible effect on peak location), we find that the three-level model reproduces the experimental binding energies' dependence on intensity for seven $\Delta_1$ values. Within the range of intensities and detuning explored in Fig.~\ref{ExpData}(a), the difference between binding energies obtained in model simulation and experiment is below \SI{4}{\kilo\hertz}.

However, deviations between experiment and model binding energies are observed at higher intensity values for small detuning (when AC Stark shift is large), as shown in Fig.~\ref{ExpData}(b). The model tends to underestimate the binding energy, with the deviation getting larger with increasing intensity. One possible reason for this is the presence of an additional, high-lying level, $\ket{Y}$, that can couple with state 1 and alter the intermediate state energy. This modifies the AC Stark shift for state 0 and state 2. We find through numerical simulations (not shown) with $\ket{Y}$ added to the three-level model [Eq.~\eqref{SimulationH}], suitable values of coupling, and $E_Y\approx 2E_1$, that the additional state can account for this deviation. Although this should not be construed as strong evidence for an additional level at this energy, it does suggest that additional levels at high energies are a candidate cause for this effect.

\begin{figure}
	\includegraphics[width=0.5\textwidth]{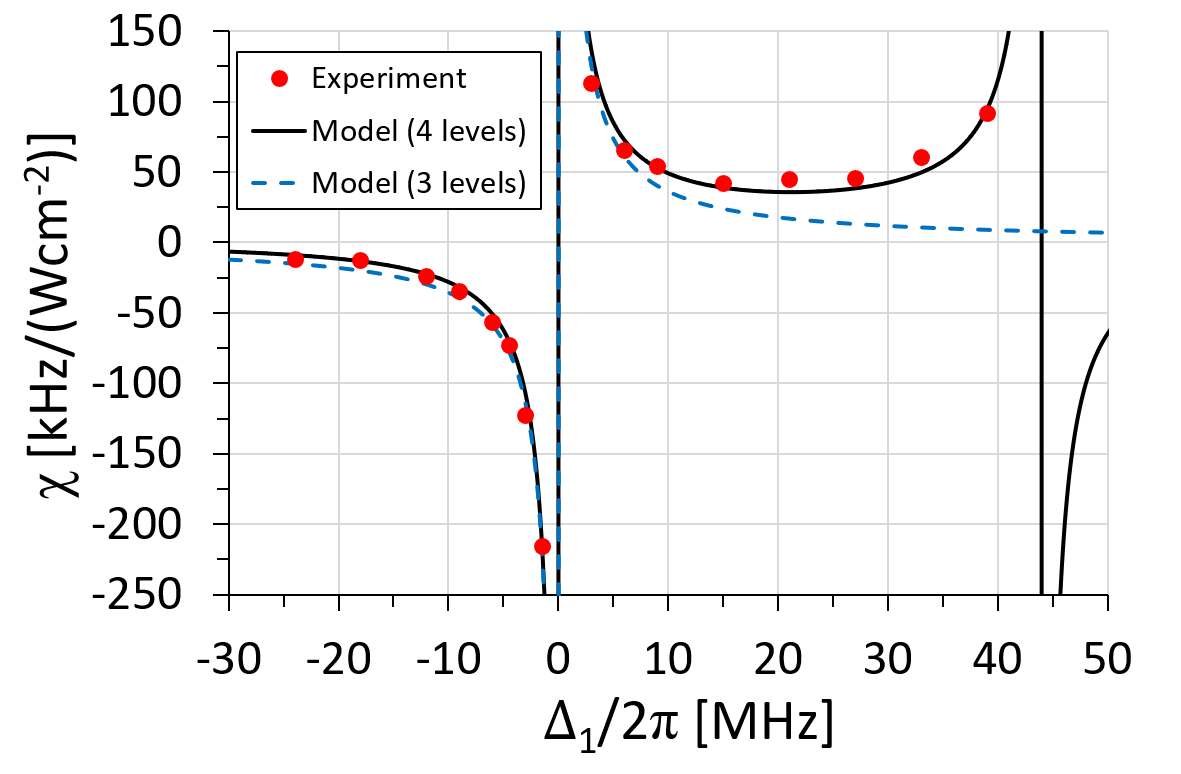}
	\caption{\label{Floquet4Level}Plot of susceptibility as a function of detuning for low-intensity data (for which light shift varies almost linearly). Note that in addition to the data from Fig.~\ref{ExpData}(a), data for detunings above \SI{6}{\mega\hertz} and below \SI{-9}{\mega\hertz} are included. The data is fitted with the susceptibility predicted from the Floquet treatment of the four-level model [First derivative of Eq.~\eqref{Floq4Eq} with respect to intensity in the low-intensity limit], with the additional state having an energy of approximately $E_1+2\pi\times$\SI{44}{\mega\hertz}. Similarly, the susceptibility obtained from the Floquet treatment of the three-level model, given by Eq.~\eqref{FloqFinalEq}, is also presented.}
\end{figure}

A second discrepancy occurs at larger intermediate state detunings ($\Delta_1$). Here, there are also significant differences between binding energy predicted by the three-level model and experimental data. Considering only low-intensity data where the light shift varies linearly with intensity, Fig.~\ref{Floquet4Level} plots the susceptibility, $\chi$, defined by $E_b-E_b^0=2\pi\chi I$, against detuning. This figure includes higher detuning data not included in Fig.~\ref{ExpData}(a). These high-detuning data show a large discrepancy, especially for data with $\Delta_1/2\pi\geq$\SI{9}{\mega\hertz}, likely due to the presence of another intermediate state close to state 1 that couples to state 0. Further analysis is provided in Sec.~\ref{Floq4Sec}. 

\subsection{Floquet Treatment}
\label{FloqMain}

To obtain analytic insight into the three-level model, we analyze the Hamiltonian with Floquet theory. Floquet theory states that for a periodic Hamiltonian $\hat{H}$ with period $T$, there exist Floquet state solutions of the form
\begin{equation}
\ket{\psi_j(t)}=e^{-i\varepsilon_jt}\ket{\phi_j(t)}
\end{equation}
where $\ket{\phi_j(t)}$ is periodic with period $T$ \cite{Shirley1965,Hanggi1998}. Substituting the Floquet state solutions into the Schr{\"o}dinger equation and performing a Fourier expansion of the Hamiltonian and $\ket{\phi_j(t)}$, we arrive at the eigenvalue equation
\begin{equation}
\sum_{k,\beta}\mel{l,\alpha}{\hat{H}_F}{k,\beta}\braket{k,\beta}{\phi_j(t)}=\varepsilon_j\braket{l,\alpha}{\phi_j(t)}
\end{equation}
where $\ket{l,\alpha}=e^{il\omega t}\ket{\alpha}$, $\omega=2\pi/T$, and $\{\ket{\alpha}\}$ is an orthonormal basis. $\hat{H}_F$ is termed the Floquet Hamiltonian and has the form $\mel{l,\alpha}{\hat{H}_F}{k,\beta} = \mel{\alpha}{H^{(l-k)}}{\beta}+l\omega\delta_{kl}\delta_{\alpha\beta}$ where $H^{(n)}$ is the n-th Fourier coefficient of the Hamiltonian $\hat{H}$. Hence, we have transformed the time-dependent periodic Hamiltonian into a time-independent one, where familiar techniques can be used to obtain solutions to the dynamics of the original Hamiltonian.

Numerical simulations indicate that shifts to the binding energy due to decay terms in the three-level model, given by Eq.~\eqref{SimulationH}, are negligible. After neglecting them, the model Hamiltonian is periodic with period $T=2\pi/\Delta\omega$. Applying Floquet theory, we obtain the Floquet Hamiltonian
\begin{equation}
\begin{split}
\hat{H}_F=&\sum_{l=-\infty}^{\infty}\biggl[-\Delta_1\dyad{l,1}{l,1}-\Delta_2\dyad{l,2}{l,2}\\
&+\frac{\Omega_{A,01}}{2}(\dyad{l,0}{l,1}+\text{h.c.})\\
&+\frac{\Omega_{B,01}}{2}(\dyad{l,0}{l-1,1}+\text{h.c.})\\
&+\frac{\Omega_{A,12}}{2}(\dyad{l,1}{l-1,2}+\text{h.c.})\\
&+\frac{\Omega_{B,12}}{2}(\dyad{l,1}{l,2}+\text{h.c.})\\
&+\sum_{\alpha=0}^{2}l\Delta\omega\dyad{l,\alpha}{l,\alpha}\biggr].
\end{split}
\end{equation}
Using the Floquet Hamiltonian, we find the binding energy following the argument of Shirley \cite{Shirley1965}. Shirley has shown that the probability of a system initially in $\ket{\alpha}$ at time $t=0$ to be in state $\ket{\beta}$ at time $t$ is
\begin{equation}
\label{ShirleyEqn}
P_{\alpha\rightarrow\beta}=\sum_l\abs{\mel{l,\beta}{e^{-i\hat{H}_Ft}}{0,\alpha}}^2.
\end{equation}
When $\Delta\omega\approx-E_b^0$, corresponding to $E_{0,2}=-\Delta_2\approx 0$ where $E_{l,\alpha}$ refers to the energy of state $\ket{l,\alpha}$, the near resonance between $\ket{0,0}$ and $\ket{0,2}$ implies a strong transition between the two states. Therefore, the term with $\ket{0,2}$ in Eq.~\eqref{ShirleyEqn} would dominate and we neglect all other $l\neq 0$ terms in the summation. Hence, we focus only on the $\ket{0,0}$ and $\ket{0,2}$ states and the near resonance condition allows us to apply second order perturbation theory in $\Omega_{A,01}$ and $\Omega_{B,12}$, with conditions $\abs{\Delta_2}\ll\{\abs{\Delta_1+l\Delta\omega}, \abs{\Delta\omega}\}$ for integer $l$, $\{\Omega_{A,01},\Omega_{B,12}\}\ll\abs{\Delta_1+j\Delta\omega}$ with $j=-1,0,1$, and $\{\abs{\Omega_{A,01}^2/4\Delta_1},\abs{\Omega_{B,12}^2/4\Delta_1}\}\ll\abs{\Delta\omega}$. The second order perturbation terms are generated via transition through the intermediate states $\ket{-1,1}$, $\ket{0,1}$ and $\ket{1,1}$, and this allows us to rewrite the Floquet Hamiltonian into a 2$\times$2 effective Hamiltonian 
\begin{equation}
\begin{split}
\mel{0,0}{\hat{H}^{\text{eff}}}{0,0}=&\frac{\Omega_{A,01}^2}{4\Delta_1}+\frac{\Omega_{B,01}^2}{4(\Delta_1+\Delta\omega)}\\
\mel{0,2}{\hat{H}^{\text{eff}}}{0,2}=&-\Delta_2+\frac{\Omega_{A,12}^2}{4(\Delta_1-\Delta_2-\Delta\omega)}\\
&+\frac{\Omega_{B,12}^2}{4(\Delta_1-\Delta_2)}\\
\mel{0,2}{\hat{H}^{\text{eff}}}{0,0}=&\frac{\Omega_{A,01}\Omega_{B,12}}{4\Delta_1}\\
\mel{0,0}{\hat{H}^{\text{eff}}}{0,2}=&\frac{\Omega_{A,01}\Omega_{B,12}}{4(\Delta_1-\Delta_2)}\approx\frac{\Omega_{A,01}\Omega_{B,12}}{4\Delta_1}.
\end{split}
\end{equation}
For $\Delta\omega\approx -E_b^0$, this analysis gives the main positive peak [peak p1 in Fig.~\ref{SystemPic}(c)]. If we use the resonance condition for the main negative peak [peak n1 in Fig.~\ref{SystemPic}(c)], $E_{-2,2}=E_b^0-\Delta\omega\approx 0$, the same expressions are obtained,  except for an additional $-2\Delta\omega$ term in $\mel{0,2}{\hat{H}^{eff}}{0,2}$. Substituting the effective Hamiltonian into Eq.~\eqref{ShirleyEqn} to obtain an expression for transition probability, we find that the transition probability amplitude at long times, and the atom loss rate, is maximized when $\mel{0,0}{\hat{H}^{\text{eff}}}{0,0}=\mel{0,2}{\hat{H}^{\text{eff}}}{0,2}$. 

\begin{figure}
	\includegraphics[width=0.5\textwidth]{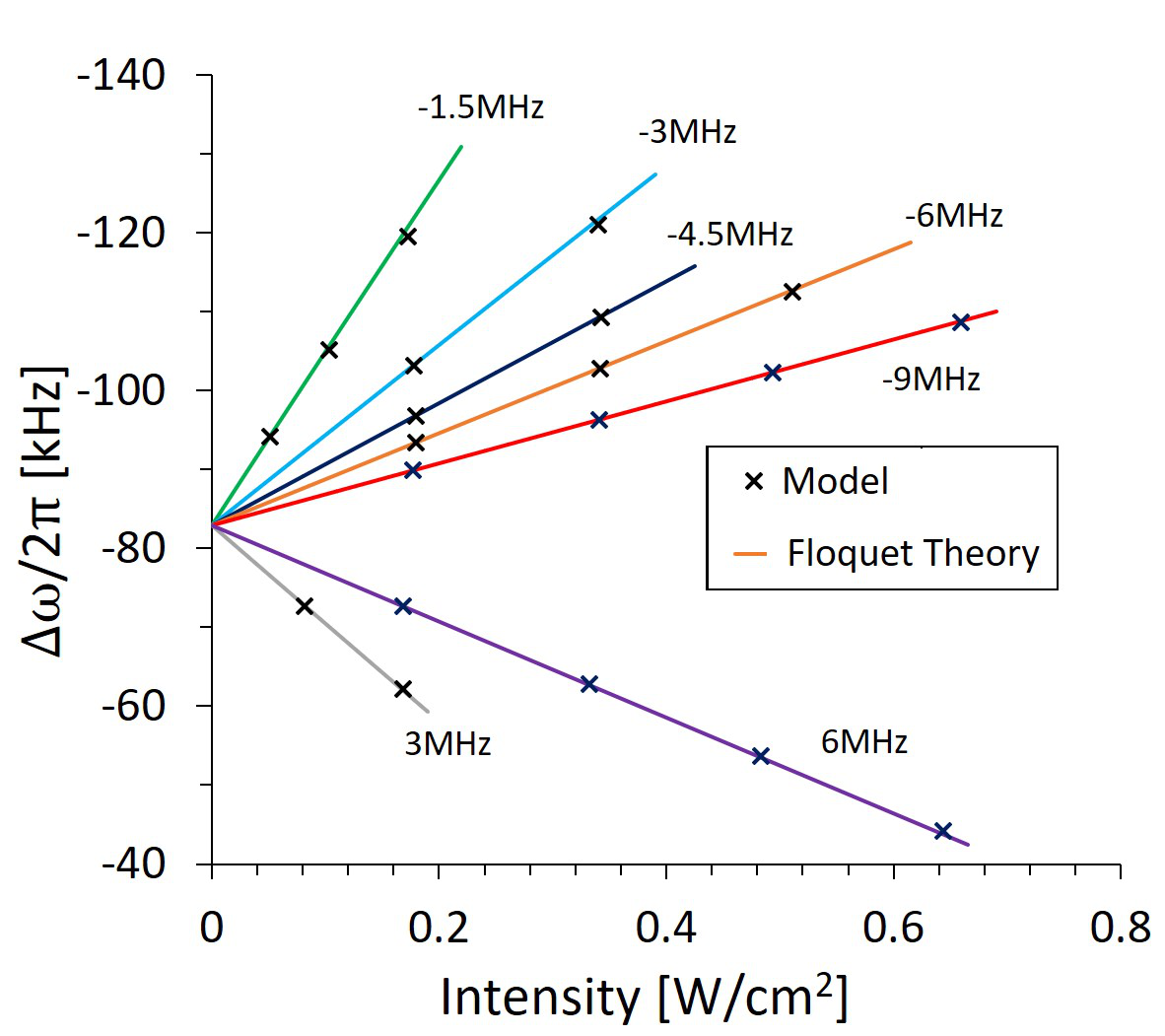}
	\caption{\label{FloquetPlot} (Color online) Binding energy as a function of intensity obtained from the full simulation, given by Eq.~\eqref{SimulationH}, and analytic solution of Floquet theory, given by Eq.~\eqref{FloqFinalEq}, at various detuning for the three-level model.}
\end{figure}

Given that the laser coupling to the $0\rightarrow1$ transition is small in the experiment ($\Omega_{A,01},\Omega_{B,01}\ll\Omega_{A,12},\Omega_{B,12}$), the binding energy can be obtained by solving the quadratic equation
\begin{equation}
\mp\Delta\omega=E_b^0+\frac{\Omega_{B,12}^2}{4(\Delta_1+E_b^0+\Delta\omega)}+\frac{\Omega_{A,12}^2}{4(\Delta_1+E_b^0)}
\end{equation}
where the negative sign corresponds to the peak associated with $\Delta\omega\approx -E_b^0$ [process p1 in Fig.~\ref{SystemPic}(b)] and the positive sign to the case of $\Delta\omega\approx E_b^0$ [process n1 in Fig.~\ref{SystemPic}(b)]. Furthermore, since the two PA lasers have the same intensity, we define $\Omega_{A,12}=\Omega_{B,12}=\Omega$ and the result simplifies to
\begin{equation}
\label{FloqFinalEq}
\mp\Delta\omega=E_b^0+\frac{\Omega^2}{4} \left(\frac{1}{\Delta_1+E_b^0+\Delta\omega}+\frac{1}{\Delta_1+E_b^0}\right).
\end{equation}

Fig.~\ref{FloquetPlot} shows that the binding energy predicted by Floquet analysis closely reproduces the results from the three-level model simulation. They closely agree even outside the approximation's strict regime of validity, since even points in Fig.~\ref{FloquetPlot} with Stark shifts of \SI{40}{\kilo\hertz} are close to the three-level model simulation results. Thus, Eq.~\eqref{FloqFinalEq} provides a simple analytic formula to analyze similar PA experiments in the future without the need to do the full three-level simulation. Eq.~\eqref{FloqFinalEq} also matches the light shift observations in past experiments~\cite{Wynar2000, Kitagawa2008}. The forms of these light shift cannot be captured by the $\Lambda$-model where only one laser drives each leg of the PA process. Upon inspection, one notices that the light shift in Eq.~\eqref{FloqFinalEq} is the sum of AC Stark shifts when the two PA lasers act independently on the $1\rightarrow 2$ transition, corresponding to the limit of $\omega_A$ and $\omega_B$ being well-separated. This suggests that for light shifts observed for a bichromatic drive, one can safely treat the effect of AC Stark shift from each laser independently at low intensities where the AC Stark shift is not too large compared to the binding energy. However, at higher intensities, the laser interference is no longer negligible and the Floquet approximation would deviate from the numerically calculated binding energy from Eq.~\eqref{SimulationH}.

\subsection{Four-Level Model}
\label{Floq4Sec}

\begin{figure}
	\includegraphics[width=0.15\textwidth]{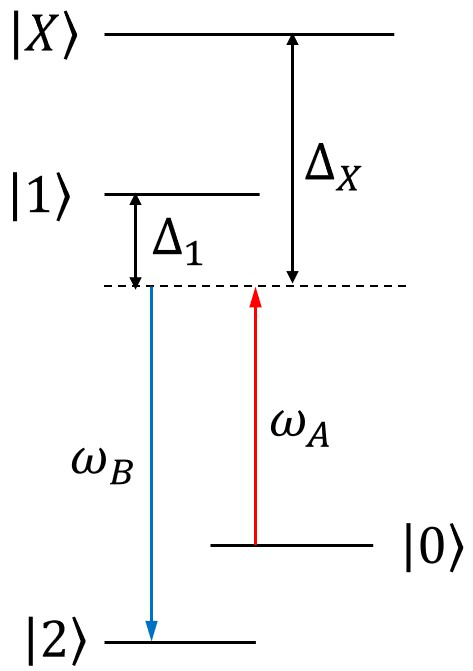}
	\caption{\label{ExtraLevel}Energy levels and couplings of the four-level model. State X is added to the three-level model and it couples to both state 0 and state 2, with laser A detuning from the $0\rightarrow X$ transition defined as $\Delta_X=\omega_A-E_X$.}
\end{figure}

Considering only low-intensity data where the binding energy varies linearly with intensity, Fig.~\ref{Floquet4Level} shows that there are deviations between susceptibility predicted by the three-level model and experimental results when the detuning is large. This may be due to the presence of another level $\ket{X}$. To test this hypothesis, we extend the model by introducing the additional level X, which couples to state 0 and is of higher energy than state 1, as shown in Fig.~\ref{ExtraLevel}. One could also assume a coupling between state 2 and X, but we will see that it is not necessary to reproduce the observations. The new Hamiltonian has three additional terms,
\begin{equation}
\begin{split}
\hat{H}'=&\hat{H}+E_X\dyad{X}\\
&+\left[\Omega_{B,0X}\cos(\omega_Bt)+\Omega_{A,0X}\cos(\omega_A t)\right]\dyad{0}{X}+\text{h.c.}
\end{split}
\end{equation}
Applying the rotating wave approximation and following the Floquet treatment, we can predict the binding energy at low intensity by solving
\begin{equation}
\label{Floq4Eq}
\begin{split}
\Delta\omega=&E_b^0+\frac{\Omega_{B,12}^2}{4(\Delta_1+E_b^0+\Delta\omega)}+\frac{\Omega_{A,12}^2}{4(\Delta_1+E_b^0)}\\
&-\frac{\Omega_{B,0X}^2}{4(\Delta_X+\Delta\omega)}-\frac{\Omega_{A,0X}^2}{4\Delta_X}
\end{split}
\end{equation}
where $\Delta_X=\omega_A-E_X$. Additionally, we introduce a parameter $k_2$ to relate the Rabi frequency of the $0\rightarrow X$ transition and laser intensity, 
\begin{equation}
\Omega_{B,0X}=2\pi k_2\sqrt{I_B}.
\end{equation}

There is close agreement between the four-level model and experimental light shifts even for large detuning values shown in Fig.~\ref{Floquet4Level}, where we performed a global fit of parameters $k$, $k_2$ and $E_X$ to susceptibility as a function of detuning. This supports the hypothesis that an additional level approximately \SI{44}{\mega\hertz} higher in energy than state 1 causes light shift deviations that are unaccounted for in the three-level model, given by Eq.~\eqref{SimulationH}. If one assumes instead that state X is coupled only to state 2, the sign of the AC Stark shift due to the additional level would be negative at red-detuning, which is inconsistent with the experimental observation in Fig.~\ref{Floquet4Level}, where the susceptibility is positive when red-detuned from state X. Hence, the dominant coupling is between state X and state 0. Two candidates for state X are unbound atoms with one atom in the excited atomic state ($^3$P$_1$+$^1$S$_0$), and the least bound $J=1$ state in the excited channel. However, for any molecular level, it is very unlikely that coupling to state 0 (free atoms) is stronger than coupling to state 2 (molecule). Therefore, we conclude that the dominant effect observed is the AC Stark shift of the atomic levels due to coupling to the $^3$P$_1$+$^1$S$_0$ collisional state, which is approximately \SI{44.2}{\mega\hertz} higher in energy than the intermediate bound state $\ket{1}$~\cite{Borkowski2014,Jim2018}.

\begin{figure}
	\includegraphics[width=0.5\textwidth]{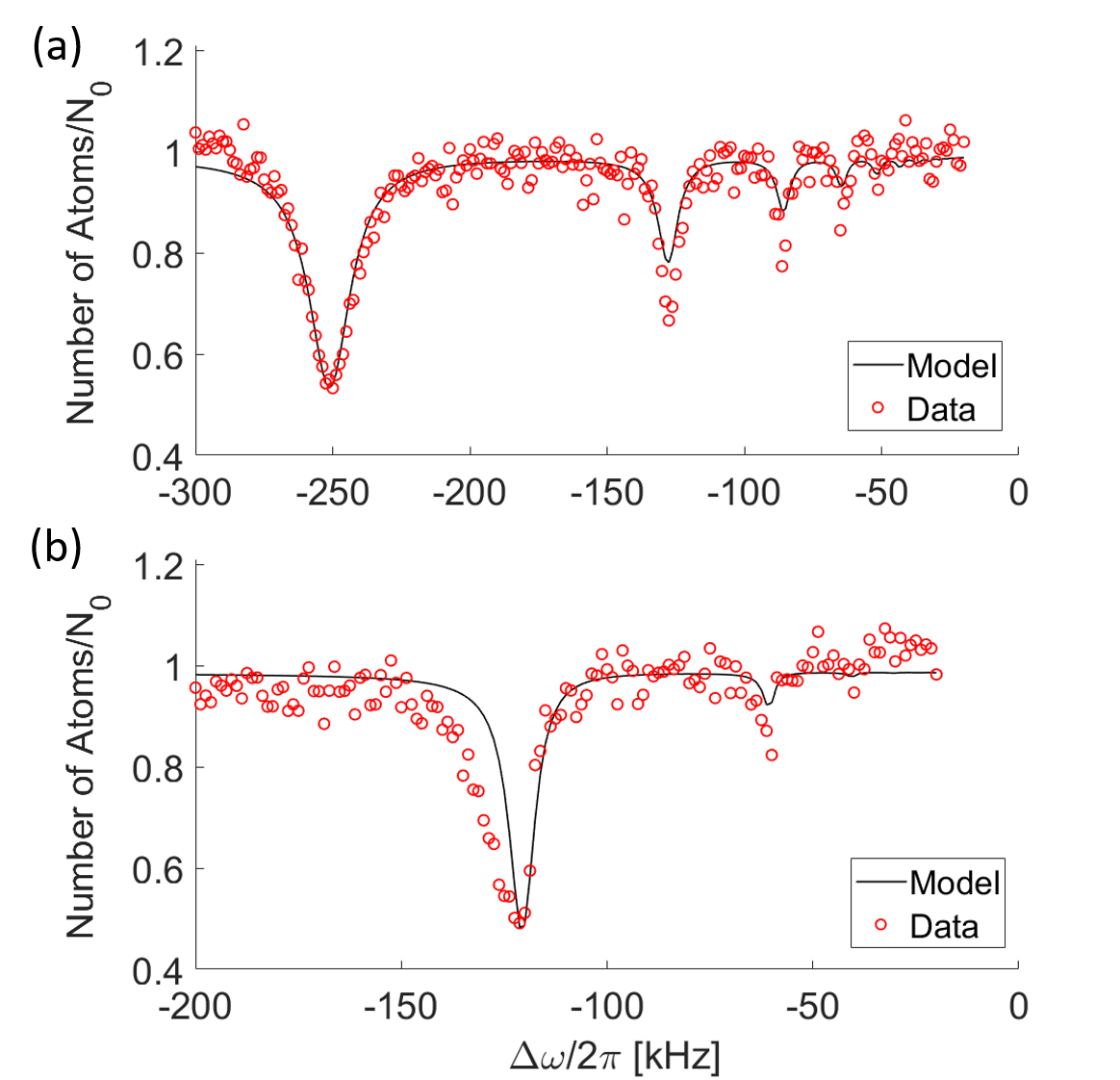}
	\caption{\label{LineShapeFit}PA spectrum obtained from experiment at $\Delta_1/2\pi=$ \SI{-1.5}{\mega\hertz} and the three-level model fits at intensities (a) $I=$ \SI{0.66}{\watt\per\centi\metre\squared} and (b) $I=$ \SI{0.17}{\watt\per\centi\metre\squared} (by varying $x$ and $\Gamma_1$). $k$ is adjusted to match the model and experimental binding energies.}
\end{figure}

\section{Lineshape}
\label{LineShapeSec}

We find that the three-level model accurately reproduces the lineshape at high laser intensities. For low-intensity spectra, the peak width and area are qualitatively accurate, but the detailed lineshape is less well reproduced, as shown in Fig.~\ref{LineShapeFit}. The model gives a symmetric lineshape, which can be adjusted to match the lineshape at high intensity where the experimental peak is symmetric. However, an asymmetry that the model is unable to capture is present in low-intensity spectral peaks. 

The asymmetry in experimental spectral peaks at low intensity results from the distribution of possible collision energies in a finite-temperature gas~\cite{Jim2018}. For an initial state consisting of two colliding $^{86}$Sr atoms, the energy of the state is given by the relative kinetic energy, $E_K$, instead of zero, where zero is the energy of two atoms at rest at large separation. This results in a larger energy gap between the initial state $\ket{0}$ and the bound state $\ket{2}$ and thus requires a larger laser frequency difference to be on resonance. The initial collisional energies ($E_K/2\pi\sim$ \SI{3}{\kilo\hertz}) should follow a Boltzmann distribution which would cause asymmetric broadening of the lineshape towards larger frequency differences~\cite{Bohn1996}. Since the model does not take into account collisional effects, it is unable to capture the asymmetry in these lineshapes accurately. At higher intensities, thermal broadening becomes less dominant compared to power broadening. Hence, we observe more symmetric lineshapes, which the model captures better.

\begin{figure*}
	\includegraphics[width=0.9\textwidth]{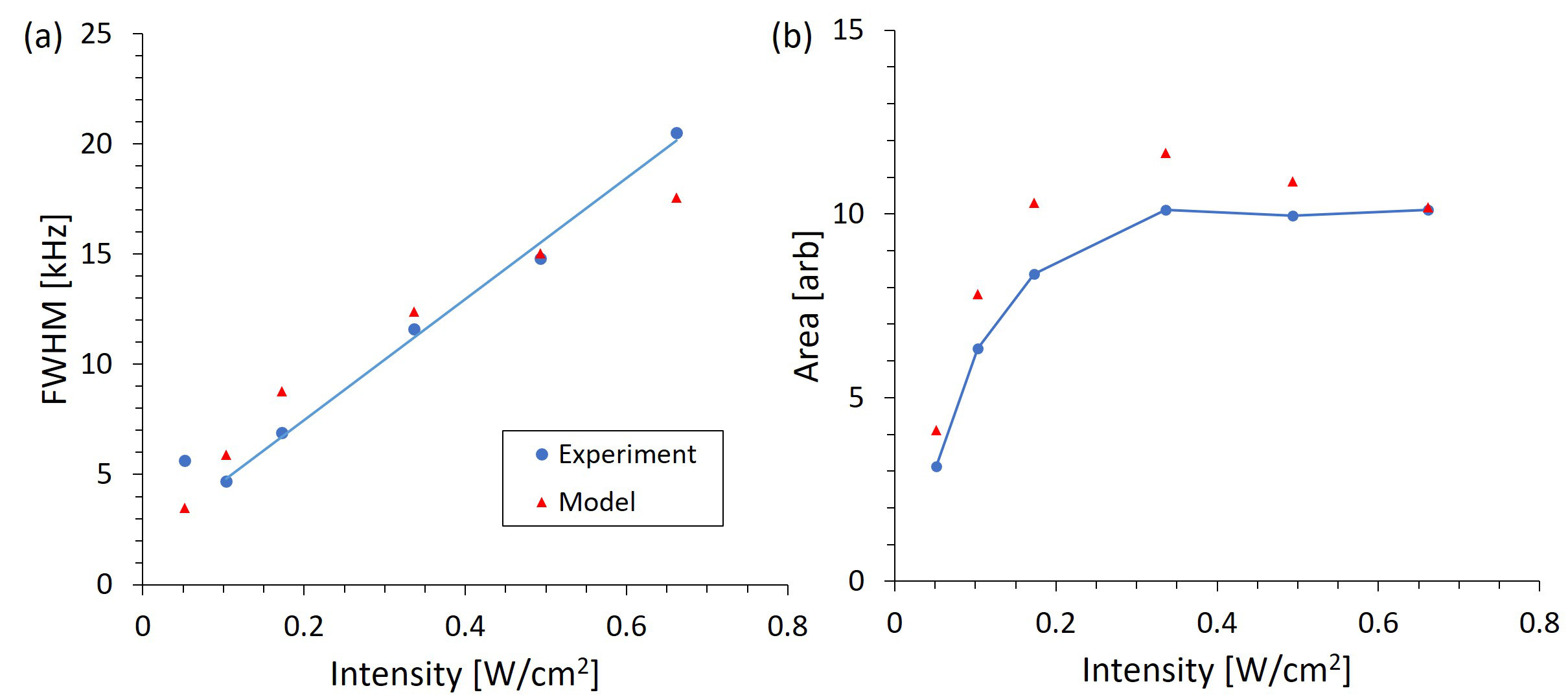}
	\caption{\label{FWHMArea}(Color online) Plot of (a) FWHM and (b) area of main negative peak at various intensities for experimental data (blue circles) and the three-level model with parameters $k=$ \SI[mode=text]{850}{kHz(W/cm^2)^{-1/2}}, $x=$ 37.5, $\Gamma_1=$ \SI{1.6e6}{\per\second} and $\Gamma_2=$ \SI{0}{\per\second} (red triangles). The line in (a) is a linear fit to the experimental data. In (b), the line serves to guide the eye through the experimental data.}
\end{figure*}

For spectra obtained at detuning $\Delta_1/2\pi=$ \SI{-1.5}{\mega\hertz}, the main negative peaks show an increasing area and line width, as measured by the full-width at half-maximum (FWHM), with intensity. As shown in Fig.~\ref{FWHMArea}(a), FWHM increases approximately linearly with intensity, suggesting that power broadening causes the increasing linewidth. The peak area shows an initial rise in value at small intensities before leveling off at a higher intensity, as shown in Fig.~\ref{FWHMArea}(b). 

The three-level model can qualitatively capture the general increase in FWHM and peak area as a function of intensity observed in experiment. Fixing $k=$ \SI[mode=text]{850}{kHz(W/cm^2)^{-1/2}} (from Sec.~\ref{SecNumFit}) and $x=$ 37.5 (to match the area of the $I=$ \SI{0.66}{\watt\per\centi\metre\squared} main negative peak, which is largely insensitive to $\Gamma_1$ and $\Gamma_2$), we test different values of $\Gamma_1$ and $\Gamma_2$, and find that the FWHM of the $\Delta_1/2\pi=$ \SI{-1.5}{\mega\hertz} main negative peaks can be reproduced over the full range of laser intensity with $\Gamma_1=$ \SI{1.6e6}{\per\second} and $\Gamma_2=$ \SI{0}{\per\second}. The plot of FWHM as a function of intensity using these parameters [red triangles in Fig.~\ref{FWHMArea}(a)] shows a general rising trend similar to experimental data, but shows a decreasing slope as intensity increases. The model also qualitatively matches the variation of area with laser intensity [Fig.~\ref{FWHMArea}(b)]. The value of $\Gamma_1$ obtained here is much higher than its expected value as discussed in Ref.~\cite{Jim2018}, whereas the value of $\Gamma_2$ is very small, consistent with its expected value. $\Gamma_1$ corresponds to the particle decay rate of state 1, the second least bound state of the $0^+_u$ molecular potential, which is expected to be \SI{9.4e4}{\per\second}, twice the decay rate of the $^3$P$_1$ excited atomic state. $\Gamma_2$ corresponds to the decay from the molecular ground state via collisions with background atoms, which is expected to be small, of the order of \SI{300}{\per\second}~\cite{Jim2018}. 

\section{Higher Order Peaks}
\label{MultiphotonSec}

\begin{table*}
	\caption{\label{MultiphotonFloq}Comparison of binding energy associated with higher order peaks, as obtained from experiment, the numerical solution of the three-level model [Eq.~\eqref{SimulationH}], the result from Floquet and perturbation theory [Eq.~\eqref{MultiphotonFloqEq}], and the naive expectation of $E_b/n$ for the n-th order peak. For the numerical and Floquet solutions, $k$ is fitted at each intensity value such that the main negative peak coincides with the experimentally determined binding energy.} 
	\begin{ruledtabular} 
		\begin{tabular}{cccccccccc} 
			Intensity & Peak 1  & \multicolumn{4}{c}{Peak 2 [\si{\kilo\hertz}]} & \multicolumn{4}{c}{Peak 3 [\si{\kilo\hertz}]}\\
			\cline{3-6} \cline{7-10}
			[\si{\watt\per\centi\metre\squared}] & [\si{\kilo\hertz}] & Exp & Num & Floq & $E_b/2$ & Exp & Num & Floq & $E_b/3$\\ 
			\hline
			0.66 & 250.4 & 127.4 & 127.4 & 128.0 & 125.2 & 86.1 & 85.5 & 86.0 & 83.5\\
			0.49 & 200.8 & 101.6 & 101.7 & 102.0 & 100.4 & 67.8 & 68.2 & 68.4 & 66.9\\
			0.34 & 161.0 & 81.6 & 81.3 & 81.4 & 80.5 & 54.1 & 54.4 & 54.5 & 53.7\\
			0.17 & 119.9 & 60.0 & 60.3 & 60.3 & 60.0 & - & - & - & -
		\end{tabular} 
	\end{ruledtabular} 
\end{table*}

Additional peaks, most prominent for spectra at high intensity and low detuning, are observed at approximately $E_b/n$ for integer $n$ in the experimental spectra shown in Fig.~\ref{SystemPic}(c-f), and these can be accounted for by the model [Eq.~\eqref{SimulationH}]. These would be absent in the $\Lambda$-model, but with the addition of laser couplings $\Omega_{B,01}$ and $\Omega_{A,12}$, the features can be understood as resulting from higher order processes. For instance, the second order resonance peaks present in Fig.~\ref{SystemPic}(c) result from a four-photon process [p2 and n2 in Fig.~\ref{SystemPic}(b)], with transitions through either 0$\xrightarrow{\omega_A}$1$\xrightarrow{\omega_B}$0$\xrightarrow{\omega_A}$1$\xrightarrow{\omega_B}$2 or 0$\xrightarrow{\omega_A}$1$\xrightarrow{\omega_B}$2$\xrightarrow{\omega_A}$1$\xrightarrow{\omega_B}$2, and detuning $E_b/2$ from states 0 or 2 in the intermediate step. These processes require either laser B driving transition $1\rightarrow 0$ or laser A driving transition $2\rightarrow 1$, which are not present in typical three-level models~\cite{Bohn1996,Yoo1985}. 

These peaks, which are absent in conventional two-frequency PA experiments, become prominent due to the unusually low binding energy provided by the halo state of $^{86}$Sr. When the binding energy is large, the detuning from states 0 and 2 in the intermediate step of the four-photon process is large, resulting in a small effective coupling between the two states relative to the coupling in the two-photon process. This results in a significantly lower transition probability through the four-photon pathway.

The experimentally observed peaks deviate slightly from the naive expectation that higher order peaks will occur at $E_b/n$, but these deviations arise from varying AC Stark shifts, and they are captured by numerical simulations of the three-level model, given by Eq.~\eqref{SimulationH}. As shown in Fig.~\ref{LineShapeFit}, the model peak locations match the experimental peaks better than the naively expected values. The Floquet treatment provides an approximate equation predicting the binding energy of the $n$-th order peak in the experiment,
\begin{equation}
\label{MultiphotonFloqEq}
n\Delta\omega=E_b^0+\frac{\Omega^2}{4}\left(\frac{1}{\Delta_1+E_b^0}+\frac{1}{\Delta_1+E_b^0+\Delta\omega}\right).
\end{equation}
The light shift varies across the full spectrum of peaks because the detuning of laser B from the $\ket{1}\rightarrow\ket{2}$ one-photon transition is varying, giving rise to a changing AC Stark shift contribution from laser B. When $\Delta\omega<0$, as in the experiment, the higher order peaks imply a higher binding energy than naively expected from the main negative peak. In contrast, when $\Delta\omega>0$, the higher order peaks imply a lower binding energy than naively expected from the main positive peak, consistent with an increasing downward shift of $\ket{2}$ with increasing frequency of laser B. For the second order perturbation theory to be valid, $\frac{\Omega^2}{4\Delta_1}\ll\Delta\omega$ (low-intensity limit). However, the experimental locations of higher order peaks relative to the main negative peak appear to be well captured even when the AC Stark shift is not low compared to the binding energy. Using Eq.~\eqref{MultiphotonFloqEq}, we fit $k$ for each intensity value such that the predicted main negative peak binding energy matches the experimental main peak, and use the same $k$ to generate predictions of binding energies of higher order peaks. We perform a similar calculation to obtain predictions from the numerical solution of the three-level model, given by Eq.~\eqref{SimulationH}. The predicted values from both the numerical solution and Floquet analysis shown in Table~\ref{MultiphotonFloq} generally match the experimental values better than the naive expectation. 

Despite reproducing the binding energy and qualitatively explaining the behavior of satellite peak areas, the three-level model does not fully capture the number of atoms lost for higher order processes, as observed in Fig.~\ref{LineShapeFit} where the higher order peaks in the three-level model simulation have a smaller area than experimentally observed. This could be linked to the model's neglect of the collisional and density-dependent aspects of the process. During the experiment, the atomic density decreases over time as atoms are lost, leading to a decrease in atomic collision rates and thus rate of atom loss, which scales as $\dv{n}{t}\propto n^2$ for a two-body scattering process~\cite{Jim2018}. In contrast, the loss rate in the model is simply proportional to the number of molecules, and thus scales as $\dv{n}{t}\propto n$. Therefore, the model overestimates the number of atoms lost when $n$ decreases and this discrepancy is most significant when the atom loss is high. Since we normalize the model results to the main negative peak, where atom loss is the highest in the spectrum, we observe that the model underestimates the atom loss in the higher order peaks, as shown in Fig.~\ref{LineShapeFit}.

\section{Conclusion}

We have shown that for bichromatic driving of a PA process in a three-level system with unperturbed binding energy that is not small compared to detuning or AC Stark shifts, it is important to consider that each laser can couple to both transition legs. Including this effect results in predictions of strong light shifts arising from both PA lasers and higher order peaks arising from multi-photon processes, which we observe in ultracold $^{86}$Sr PA experiments. 

Using a simple three-level model, we explain the origin of higher order peaks as alternate resonance processes that only appear in the simulations when the additional laser couplings, $\Omega_{A,12}$ and $\Omega_{B,01}$, are included. The model also predicts the binding energies at various detuning and intensities with predominantly two fitting parameters, $k$ and $E_b^0$, and weak dependence on $x$, $\Gamma_1$, and $\Gamma_2$. We derive simple analytic forms for the binding energy by combining Floquet and perturbation theory, confirming earlier conjectures that the light shift is a sum of independent AC Stark shifts from each of the lasers~\cite{Kitagawa2008,Wynar2000}. Although the analytic results rely on certain limits, the numerical calculations and comparison with experiment indicate that these expressions are quite accurate for most intensities and detunings studied in this work. This provides a foundation for future work, circumventing the need for simulations when analyzing experimental PA spectra under many experimentally-relevant conditions.

We also consider the lineshapes. The model is able to capture the symmetric lineshapes for high-intensity spectral peaks, and also replicate qualitatively the general trends in linewidth and area for \SI{-1.5}{\mega\hertz} detuning data. However, the model is unable to reproduce the asymmetric broadening in low-intensity spectral peaks, likely due to neglecting collisional and scattering effects. Therefore, to have a complete theoretical model of the PA spectral lineshapes, it is essential to include the collisional effects of two-atom scattering. 

The presence of a weakly bound halo state of $^{86}$Sr$_2$ at approximately \SI{-83.0}{\kilo\hertz}, determined in Ref.~\cite{Jim2018}, and the strong light shift experienced by this state, observed in this work, open up the possibility of using coupling to the intermediate state, $\ket{1}$, to tune the halo state into collisional resonance, and thus tune the atom-atom scattering length. This could serve as a mechanism for optically controlling a Feshbach resonance in alkaline-earth atoms, where a magnetic Feshbach resonance is not possible due to the lack of hyperfine structure. The ability to easily tune the scattering length broadens the explorable range of few and many-body physics. Our results here, both theory and experimental, provide an foundation to implement this technique.

\begin{acknowledgments}
	This work was supported by the Welch Foundation (C-1844 and C-1872) and the National Science Foundation (PHY-1607665). W. Y. K. thanks Nanyang Technological University for financial support through the CN Yang Scholars Programme.
\end{acknowledgments}

\bibliography{Sr-PARef}

\end{document}